\documentclass[12pt]{article}
\usepackage{epsf}
\usepackage{amssymb}
\def\refw{\smallskip\hangindent=1.5pc\hangafter=1\noindent}
\begin{document}

\noindent
{\bf PHOTOMETRIC OBSERVATIONS OF THE SUPERNOVA 2009nr}

\medskip
\noindent
{\bf D. Yu. Tsvetkov$^1$, P. V. Balanutsa$^1$,
V. M. Lipunov$^1$, I. M. Volkov$^1$, 
O. A. Tuchin$^2$, 
I. P. Kudelina$^3$, M. V. Pruzhinskaya$^1$,
E. S. Gor\-bov\-skoy$^1$, V.G.Kornilov$^1$, 
A. A. Belinskii$^1$, N. V.~Tyurina$^1$, 
V. V.~Yurkov$^3$, Yu. P.~Sergienko$^3$,
A. G.~Tlatov$^4$, 
A. V.~ Parkhomenko$^4$, D. V.~Dor\-mi\-don\-tov$^4$,
V. A.~Senik$^4$, and V. V. Krushinskii$^5$
} 

\medskip
\noindent
{\it
$^1$Sternberg Astronomical Institute, Lomonosov Moscow State University,
Universitetskii pr. 13, Moscow, 119992 Russia\\ 
$^2$Children's Additional Education Center of Industrial 
District, ul. Novo-Vokzal'naya 203a, Samara, Russia\\
$^3$Blagoveshchensk State Pedagogical University, ul. Lenina 104, 
Bla\-go\-vesh\-chensk, 675000 Russia\\ 
$^4$Kislovodsk Solar Station, Main Astronomical Observatory, 
Russian Academy of Sciences, 
Kislovodsk, 357700 Russia\\ 
$^5$Ural State University, ul. Mira 19, Yekaterinburg, Russia
} 

\bigskip
\noindent
{\bf Abstract} -- We present the results of our $UBVRI$  CCD photometry 
for the second brightest supernova 
of 2009, SN 2009nr, discovered during a sky survey with the 
telescopes of the MASTER robotic network. 
Its light and color curves and bolometric light curves have been constructed. 
The light-curve parameters 
and the maximum luminosity have been determined. 
SN 2009nr is shown to be similar in light-curve 
shape and maximum luminosity to SN 1991T, which is the prototype 
of the class of supernovae Ia with an 
enhanced luminosity. SN 2009nr exploded far from the center of 
the spiral galaxy UGC 8255 and most 
likely belongs to its old halo population. 
We hypothesize that this explosion is a consequence of the merger 
of white dwarfs. 

\medskip
\noindent
DOI: 10.1134/S1063773711110053 


\medskip
{\bf INTRODUCTION} 

\medskip
SN 2009nr was discovered in the CCD images 
obtained with the telescope of the MASTER robotic 
network (Lipunov et al. 2010) in Blagoveshchensk 
on December 22, 2009; the report on the discovery 
was published on January 6, 2010 (Balanutsa and 
Lipunov 2010). The SN brightness at the time 
of its discovery was about $13^m.6$, which made it 
the second brightest SN discovered in 2009 (after 
SN 2009ig). On January 6, 2010, SN 2009nr was 
independently discovered during observations as part 
of the supernova search program at the Lick Observatory
(Li et al. 2010). According to Balanutsa and 
Lipunov (2010), the coordinates of SN 2009nr are 
$\alpha =13^{\rm h}10^{\rm m}58^{\rm s}.95;  \delta =+11^{\circ}29'29''.3$ 
(J2000.0); 
Li et al. (2010) give the last digits $58^{\rm s}.94$  and $29''.6$. 
The SN was $36''$ east and $50''$ 
north of the center 
of the Scd galaxy UGC 8255. Foley and Esquerdo 
(2010) reported that the spectrum of SN 2009nr was 
taken on January 7, 2010, with the 1.5-m Whipple 
Observatory telescope; it showed that the object 
belongs to the class of SNe Ia similar to SN 1991T. 
SN 2009nr was found in the ASAS images, which 
made it possible to trace its light curve near the 
maximum (Khan et al. 2011). 

\begin{figure}
\begin{center}
\epsfxsize=12cm
\epsffile{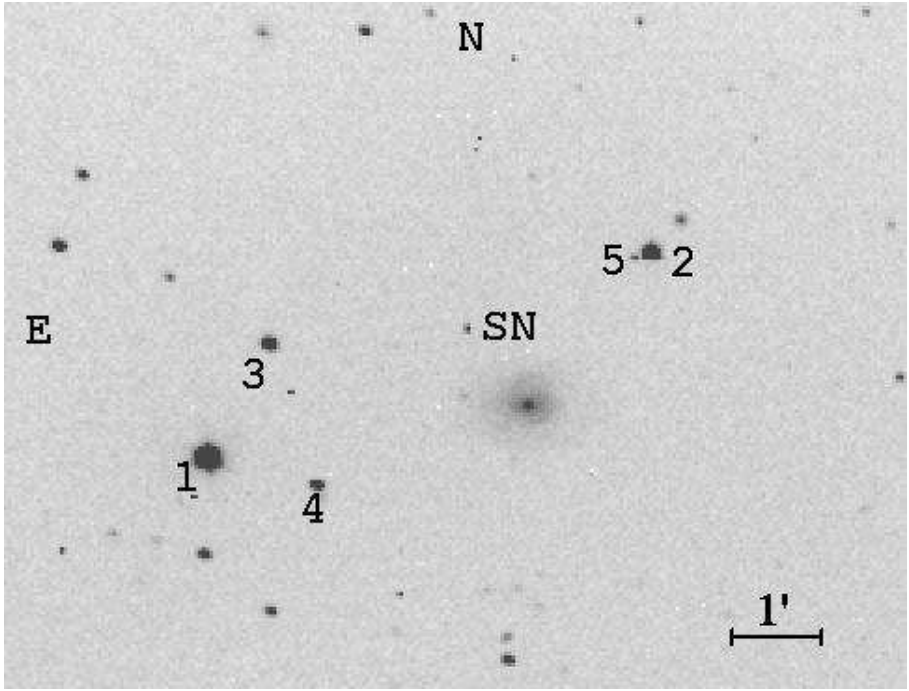}
\end{center}
\caption{SN 2009nr and the comparison stars. The image 
was obtained at S50 with the $R$  filter}
\end{figure} 

\medskip
{\bf OBSERVATIONS AND REDUCTIONS} 

\medskip
We performed our photometric CCD observations
of SN 2009nr with the following instruments 
(their abbreviated designations are given in parentheses):
the telescopes of the MASTER robotic 
network in Kislovodsk with $V,R$  filters (MK) and 
in Blagoveshchensk without filters (MB) (Lipunov 
et al. 2010); the 1-m and 60-cm telescopes of the 
Simeiz Observatory (Crimean Astrophysical Observatory)
with a VersArray 512FUV camera and $B,V,R$ 
filters (C100, C60); the 50-cm telescope of 
the Astronomical Institute of the Slovak Academy of 
Sciences in Tatranska Lomnica with an ST-10XME 
CCD camera and $U,B,V,R,I$  filters (S50); the 
20-cm meniscus telescope of the Sternberg Institute 
in Moscow with an AP-7p camera in the $B,V,R,I$  
bands (M20); and the 2-m Faulkes Telescope North 
(FTN) with $u,B,V,R,i$  filters. 

We measured the SN brightness relative to the 
comparison stars by the method of PSF photometry 
in the IRAF\footnote[1]{IRAF is distributed by the NOAO, which are operated 
by the AURA, Inc., under contract with the NSF}
DAOPHOT package. The galaxy's 
background near the SN is negligible and does not 
affect the accuracy of our photometry. The comparison
stars are shown in Fig. 1; their magnitudes 
are given in Table 1. The brighter stars 1 and 2 
were calibrated using standards from Landolt (1992) 
and standard stars in the cluster M67 (Chevalier and 
Ilovaisky 1991) on the images obtained in March 
2010 with M20 and in March 2011 with the 70-cm 
Sternberg Astronomical Institute reflector. For the 
remaining stars, we used SDSS\footnote[2]{http://www.sdss.org}
photometric data 
reduced to the $UBVRI$  system by means of relations 
from Chonis and Gaskell (2008). These magnitudes 
are in good agreement with the calibration results for 
stars 3 and 4 from our observations, which, however, 
have a fairly low accuracy. 

\begin{table}
\caption{Magnitudes of local standard stars}\vskip2mm
\begin{tabular}{ccccccccccc} 
\hline
Star & $U$ & $\sigma_U$ & $B$ & $\sigma_B$ & $V$ & $\sigma_V$ & $R$ & 
$\sigma_R$ &
$I$ & $\sigma_I$ \\
\hline
1 & 13.42& 0.03& 12.80& 0.02&  11.88& 0.01&  11.38& 0.01&  10.91& 0.01\\  
2 & 14.68& 0.04& 14.36& 0.03&  13.58& 0.02&  13.15& 0.01&  12.73& 0.02\\  
3 & 15.75& 0.05& 15.58& 0.02&  14.89& 0.02&  14.49& 0.02&  14.12& 0.02\\  
4 & 16.67& 0.05& 16.49& 0.03&  15.74& 0.03&  15.32& 0.03&  14.94& 0.03\\  
5 & 20.45& 0.08& 19.28& 0.05&  17.78& 0.05&  16.87& 0.05&  16.00& 0.05\\
\hline
\end{tabular}
\end{table}

The equations for the reduction of instrumental 
magnitudes to standard ones were derived for C100, 
C60, M20, and S50 from observations of the cluster
M67. Since the same CCD camera and set of 
filters were used for C100 and C60, the reduction 
coefficients are identical. The coefficients do not differ
greatly from those in Tsvetkov et al. (2008) and 
Elmhamdi et al. (2011): they are rather small for 
the $V$  and $R$  bands; for the $B$  band, they are $K_B = 
-0.15, -0.12, 0.15$  for C100, M20, and S50, respectively;
for the $I$  band, they are $K_I = -0.5, -0.04$ 
for M20 and S50. Since we have no data for the 
FTN and MASTER that would allow the reduction 
equations to be established, we applied no corrections 
for the possible deviations of the instrumental system 
from the standard one. The $V$  and $R$  filters at the 
MASTER telescopes and the $B,V,R$  filters at the 
FTN realize the standard Johnson-Cousins system 
(Bessell 1990); therefore, the coefficients must be 
small. The $u$  and $i$  filters realizing the SDSS system 
are used at the FTN, but we, nevertheless, calibrated 
the images in these filters by the $U$  and $I$  magnitudes.
 This should not introduce significant errors 
for the $U$  magnitudes, because the response curve for 
the SDSS $u$  band is close to the standard $U$  band. 
However, systematic errors for the $I$  magnitudes are 
possible; accordingly, we increased the errors in the 
magnitudes in Table 2. 

\begin{table}
\caption{Observations of SN 2009nr}\vskip2mm
\begin{tabular}{cccccccccccl}
\hline
JD 2455000+ & $U$ & $\sigma_U$ & 
$B$ & $\sigma_B$ & $V$ & $\sigma_V$ & $R$ & $\sigma_R$ &
$I$ & $\sigma_I$ & Tel.\\
\hline
188.40 &      &     &      &      & 13.99 & 0.04&      &      & & &MB\\
188.43 &      &     &      &      & 13.99 & 0.04&      &      & & &MB\\
196.58 &      &     &      &      & 13.80 & 0.04& 13.73& 0.03 & & &MK\\
196.61 &      &     &      &      & 13.79 & 0.03& 13.71& 0.04 & & &MK\\
197.47 &      &     &      &      & 13.82 & 0.03& 13.74& 0.04 & & &MK\\
197.48 &      &     &      &      & 13.80 & 0.13& 13.72& 0.03 & & &MK\\
198.54 &      &     &      &      & 13.80 & 0.03& 13.74& 0.03 & & &MK\\
201.51 &      &     &      &      & 13.90 & 0.03& 13.85& 0.03 & & &MK\\
203.61 &      &     &      &      & 13.96 & 0.04& 13.88& 0.12 & & &MK\\
208.09 & 14.96& 0.07& 14.90& 0.05 & 14.25 &0.07 & 14.24& 0.05 &      &      & FTN  \\
213.70 & 15.62& 0.10& 15.46& 0.03 & 14.61 &0.03 & 14.49& 0.03 & 14.51& 0.03 & S50 \\
216.59 &      &     &      &      & 14.87 &0.05 & 14.46& 0.04 &    &  & MK\\
219.07 & 16.17& 0.09& 15.94& 0.05 & 14.81 &0.05 & 14.53& 0.05 & 14.58& 0.08 & FTN \\
219.69 & 16.09& 0.08& 15.92& 0.04 & 14.86 &0.03 & 14.63& 0.03 & 14.56& 0.03 & S50 \\
234.45 &      &     & 16.56& 0.19 & 15.55 &0.15 & 15.30& 0.07 & 15.01& 0.17 & M20\\
236.55 &      &     &      &      & 15.83 &0.09 & 15.30& 0.08 &  & & MK\\
238.55 &      &     &      &      & 15.83 &0.10 & 15.46& 0.08 &  & & MK\\
244.47 &      &     & 16.93& 0.20 & 15.99 &0.12 & 15.66& 0.11 &      &      & M20\\
253.11 & 17.35& 0.07& 17.03& 0.04 & 16.20 &0.05 & 15.98& 0.05 & 16.09& 0.08 & FTN\\
259.61 &      &     & 17.10& 0.06 & 16.37 &0.03 & 16.26& 0.03 &      &      & C100\\
266.44 &      &     & 17.20& 0.03 & 16.53 &0.02 & 16.48& 0.02 &      &      & C100\\
267.42 &      &     &      &      &       &     & 16.69& 0.14 &      &      & M20\\
279.57 &      &     & 17.45& 0.05 & 16.89 &0.04 & 16.84& 0.04 & 17.19& 0.07 & S50\\
305.40 &      &     & 17.78& 0.06 & 17.54 &0.03 & 17.69& 0.03 &      &      & C60\\
309.43 &      &     &      &      & 17.44 &0.17 & 17.58& 0.19 &      &      & C60\\
315.43 &      &     &      &      &       &     & 17.92& 0.10 &      &      & C60\\
317.41 &      &     &      &      &       &     & 17.95& 0.10 &      &      & C60\\
323.97 & 18.81& 0.08& 18.14& 0.05 & 17.73 &0.07 & 18.08& 0.05 & 18.36& 0.12 & FTN\\
344.38 &      &     &      &      & 18.26 &0.07 & 18.62& 0.08 &      &      & C60\\
345.29 &      &     & 18.26& 0.08 & 18.26 &0.06 & 18.55& 0.06 &      &      & C60\\
352.31 &      &     & 18.37& 0.07 & 18.44 &0.05 & 18.93& 0.11 &      &      & C60\\
357.39 &      &     &      &      & 18.42 &0.07 & 19.05& 0.12 &      &      & C60\\
369.36 &      &     &      &      &       &     & 19.43& 0.25 &      &      & C60\\
\hline
\multicolumn{12}{l}{{\sl Note:} the errors of the published paper are corrected in 
this version of Table 2}\\
\end{tabular}
\end{table}

The MASTER images without a filter in which 
SN 2009nr was discovered were calibrated by the 
$V$  magnitudes. The subsequently obtained images 
without a filter were not reduced, because there were 
much data obtained with filters in this period. 

The results of our SN brightness measurements 
are presented in Table 2. 

\begin{figure} 
\begin{center} 
\epsfxsize=12cm
\epsffile{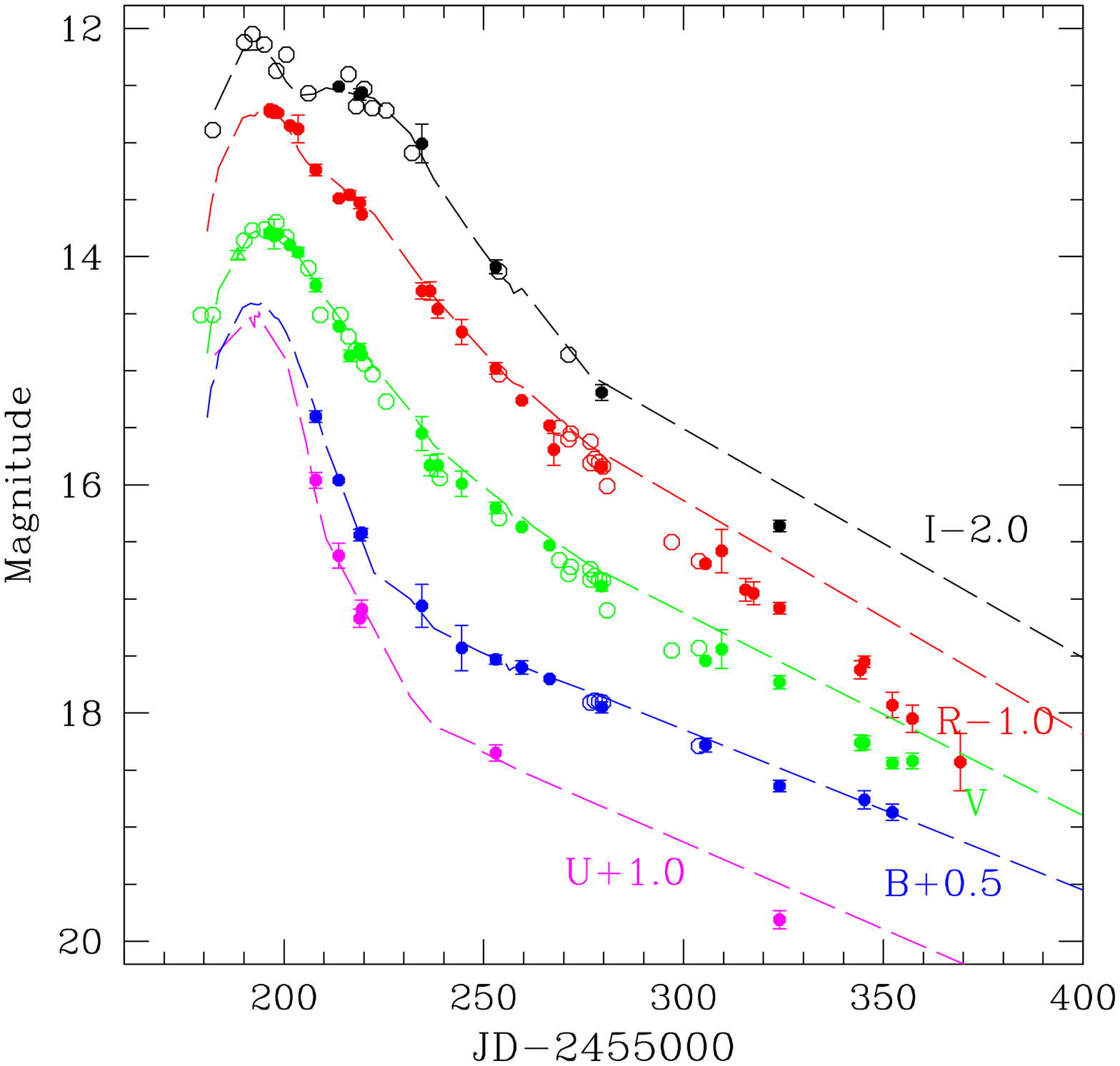}
\end{center}
\caption{Light curves of SN 2009nr. Our data obtained with and 
without filters are indicated by the filled circles and triangles, 
respectively. The open circles represent the data from 
Khan et al. (2010). The dashed lines are the light curves of SN 1991T}
\end{figure} 

\begin{figure} 
\begin{center} 
\epsfxsize=12cm
\epsffile{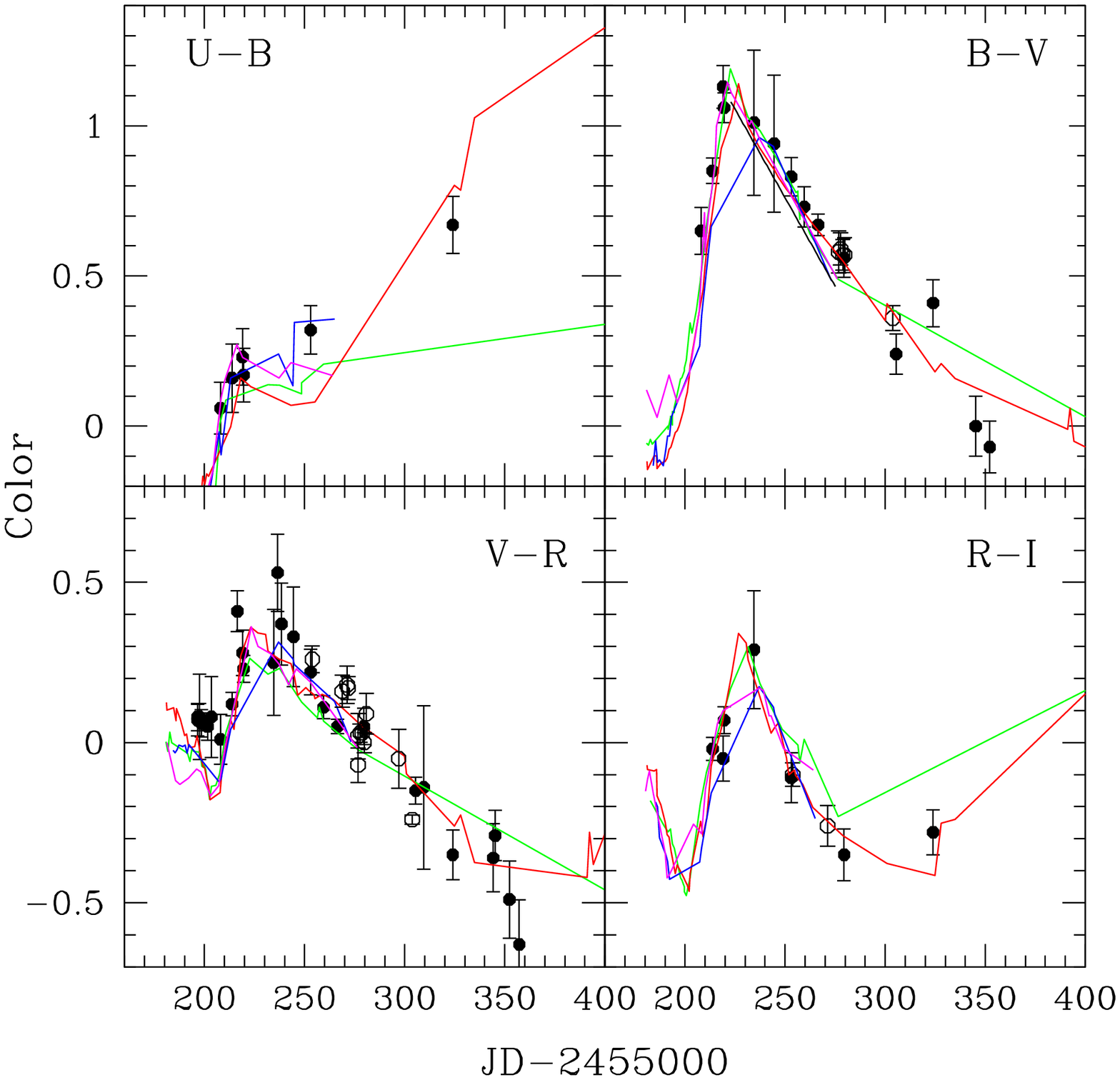}
\end{center}
\caption{Color curves of SN 2009nr. Our data and the data from 
Khan et al. (2010) are indicated by the filled and open circles, 
respectively. The color curves of several supernovae are 
indicated by the lines: SN 1991T (green), SN 1998es (blue)
SN 2003du (red), SN 1999aa (magenta). The black line on 
the $B-V$  diagram is the Lira-Phillips relation}
\end{figure}

\medskip
{\bf RESULTS OF OUR OBSERVATIONS}
 
\medskip
The light curves are shown in Fig. 2. We present 
the results from Khan et al. (2011), which are in 
good agreement with our magnitudes in all filters, 
along with other data. The light curves of SN 1991T 
(Lira et al. 1998), which belongs to the same SN Ia 
subclass as SN 2009nr, are shown for comparison.
The points of maximum light in the $V$ and $I$  
bands for SN 2009nr can be found directly from the 
observational data: $V_{max} =13.73\pm0.08, I_{max}=14.07\pm0.06, 
t_{Vmax}={\rm JD}2455194\pm1, t_{Imax}={\rm JD}2455192\pm1$. 
The light curves of SN 1991T near 
the maximum are in good agreement with those of 
SN 2009nr; the difference begins 20-30 days after 
the maximum, when the decline rate of SN 2009nr 
becomes higher in all bands except $B$. The parameter 
$\Delta m_{15}$  (the brightness decrease in 15 days after the 
maximum) for the $V$  light curve can be directly 
determined; it is $0.59\pm 0.05$. For SN 1991T and 
for SNe 1998es and 1999aa, which also belong to 
the SN Ia 1991T subclass (Hicken et al. 2009; 
Jha et al. 2006; Tsvetkov and Pavlyuk 2004), the 
values of $\Delta m_{15}(V)$  are close to this value (being, 
respectively, 0.62, 0.59, and 0.58). For these three 
SNe, $\Delta m_{15}(B)$  is approximately the same, about 0.8 
(Hicken et al. 2009); it can be assumed that its 
value is close that for SN 2009nr. The decline 
rates in the phase interval 100-160 days after the 
maximum in the $B,V,R$  bands are, respectively, 
1.4, 2.0, 2.6 magnitudes in 100 days. Comparison 
with SN 1991T and the "normal" SN Ia 2003du 
(Stanishev et al. 2007) shows that the decline rate in 
this phase interval in the $B$  band is approximately the 
same for the three objects, but it differs in the $V$  and 
$R$  bands, with the lowest and highest values being 
observed for SN 2003du and SN 2009nr, respectively. 

The color curves of SN 2009nr are shown in Fig. 3, 
where they are compared with those for SNe 1991T, 
1998es, 1999aa, and 2003du. The color curves of 
SN 2009nr, 1999aa, and 2003du were corrected for 
the extinction in our Galaxy, respectively, 
$E(B-V)_{Gal} =0.03, 0.04, 0.01$  (Schlegel et al. 1998), while 
the color curves of SN 1991T and 1998es were corrected
for the total extinction assumed to be the 
same for them, $E(B-V)_{tot} =0.15$  (Lira et al. 1998; 
Tsvetkov and Pavlyuk 2004). Figure 3 also plots the 
Lira-Phillips relation (Phillips et al. 1999) showing 
the time dependence of $(B-V)$  in the phase interval 
30-90 days for most of the SNe Ia that suffered no 
extinction. The color curves of SN 2009nr exhibit no 
noticeable reddening, which confirms the conclusion 
by Khan et al. (2010) about the absence of extinction 
in the host galaxy. Note that the $(U-B)$, $(V-R)$, and 
$(R-I)$  color curves of SN 2009nr differ from those 
for SN 1991T and show the greatest similarity to the 
color curves of the “normal” SN Ia 2003du. 

For the distance modulus of the galaxy UGC 8255, 
we take the value obtained from its radial velocity 
corrected for the velocity field of the Virgo cluster
neighborhood with the Hubble constant $H_0=73$  km s$^{-1}$Mpc$^{−1}$, 
$\mu$ =33.44\footnote[3]{http://nedwww.ipac.caltech.edu/}.
The absolute 
magnitude of SN 2009nr at maximum light is then 
$M_V=-19.8$. This is a very high luminosity for 
SNe Ia, even for the SN 1991T subclass distinguished
by an enhanced luminosity. The light curve 
of SN 2009nr in absolute $V$  magnitudes is shown 
in Fig. 4, where the light curves of SNe 1991T, 
1998es, 1999aa, and 2003du are plotted for comparison.
The color excesses adopted for these supernovae
are given above; the extinction was calculated
at $R_V =3.1$. The distance moduli taken for 
SNe 1998es, 1999aa, and 2003du are, respectively, 
$\mu$ =33.19, 33.89, 32.50; they were calculated in the 
same way as for UGC 8255 and were obtained from 
the same source; the distance modulus for SN 1991T, 
$\mu$ =30.76, was determined from observations of 
Cepheids (Saha et al. 2006). Figure 4 also shows 
the "quasi-bolometric" light curves of the same SNe 
obtained by integrating the flux in the range from $U$ 
to $I$. The bolometric light curves and the light curves 
in absolute $V$  magnitudes for SNe 2009nr and 1991T 
nearly coincide up to a phase of 80 days; the difference is 
observed only at later stages. SN 1998es is slightly 
brighter than SN 2009nr, while SNe 1999aa and 
2003du are considerably fainter. 

\begin{figure} 
\begin{center} 
\epsfxsize=12cm
\epsffile{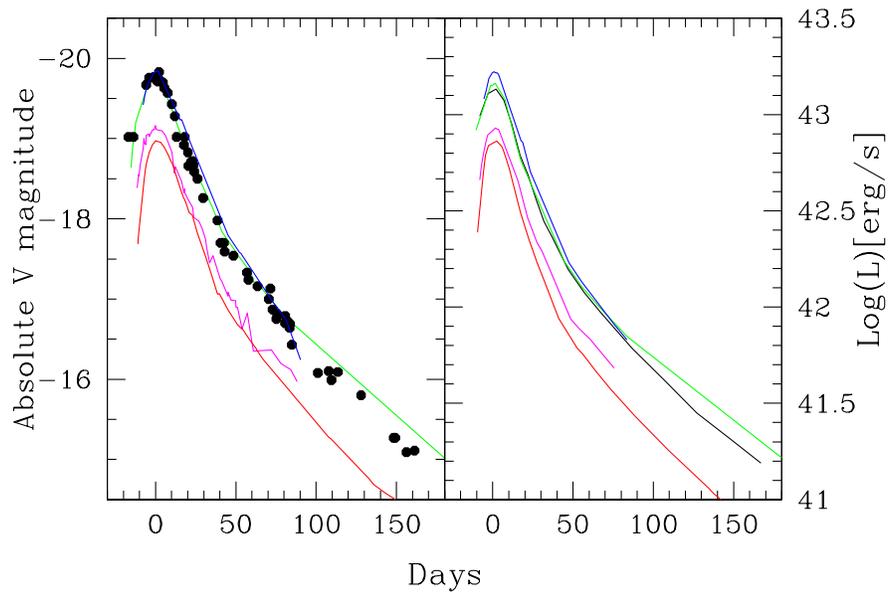}
\end{center}
\caption{Light curve of SN 2009nr in absolute $V$  magnitudes 
(left panel, filled circles) and “quasi-bolometric” light curve (right 
panel, black line). 
For comparison, the light curves of SN 1991T, 1998es, 1999aa, and 2003du 
are shown (by the same lines as 
those in Fig. 3). The time is given in days since the maximum 
in the $V$  band}
\end{figure}

According to "Arnett’s rule" (Arnett 1982), the 
luminosity of SNe Ia at maximum light is determined 
by the rate of energy release during radioactive $^{56}$Ni 
decay. The simple equations proposed, for instance, 
by Stritzinger et al. (2006) and Hayden et al. (2010) 
relate the mass of $^{56}$Ni synthesized during the explosion
to the bolometric luminosity (or absolute $V$  magnitude)
and the time interval from the explosion to 
maximum light. Assuming that the maximum light 
of SN 2009nr in the $V$  band occurred 19.7 days after 
the explosion (Khan et al. 2011), we find $M(^{56}{\rm Ni})= 
0.73M_{\odot}$
from the formula for the bolometric luminosity
 and $M(^{56}{\rm Ni})=1.07M_{\odot}$
from the formula for 
the absolute $V$  magnitude. The difference between 
these estimates is partly explained by the fact that 
the “quasi-bolometric” luminosity we derived disregards
the ultraviolet and infrared radiation; the $^{56}$Ni 
mass estimated in this way should be increased by 
10-20\%. We obtain the following $^{56}$Ni mass estimates
for SNe 1991T, 1998es, 1999aa, and 2003du 
based on their absolute magnitudes and bolometric 
luminosities by assuming the mean time of maximum 
light to be 19 days: 1.00, 0.73; 1.08, 0.89; 0.58, 
0.43; 0.48, 0.37. SN 2009nr shows a similarity to 
SNe 1991T and 1998es, while the $^{56}$Ni mass estimates
for SNe 1999aa and 2003du are almost a factor 
of 2 lower. It can be noted that, by its spectroscopic 
characteristics, SN 1999aa is believed to be an intermediate
object between the "normal" SNe Ia and the 
SN 1991T-type SNe (Garavini et al. 2004). 

Several software packages have been developed 
in recent years to determine the light-curve parameters
for SNe Ia (see, e.g., Kessler et al. 2009b; 
Conley et al. 2008; Burns et al. 2011). We reduced
our observations of SN 2009nr using the 
SNOOPY code (Burns et al. 2011) based on the 
light-curve standardization technique developed by 
Prieto et al. (2006). Two versions of the code 
were applied. The first allows one to determine the 
maximum light in the $B,V,R,I$  bands, the time of 
maximum in the $B$  band $t_{Bmax}$, and the parameter 
$\Delta m_{15}$  (it corresponds to $\Delta m_{15}(B)$ 
but is determined 
by the light-curve fitting in all filters); the second 
provides estimates of the extinction in the host galaxy 
$E(B-V)_{host}$, its distance modulus $\mu$, $t_{Bmax}$,  and 
$\Delta m_{15}$. 

The following results were obtained. 

The first version: $B_{max} =13.66\pm 0.02, V_{max} = 
13.66 \pm 0.02, R_{max} =13.64 \pm 0.02, I_{max} = 
14.08 \pm 0.02, t_{Bmax} = {\rm JD}2455192.3 \pm 0.3, 
\Delta m_{15} = 0.96 \pm 0.02.$ 
The second version: $E(B-V)_{host} =0.11 \pm 0.03, 
t_{Bmax} = {\rm JD}2455192.0 \pm 0.2, \Delta m_{15} =0.82 \pm 0.03, 
\mu =32.93 \pm 0.04$. 

The estimates of the maximum light in the $V$  and 
$I$  bands obtained in this way virtually coincide with 
those determined directly from observational data; 
the code also makes it possible to determine $B_{max}$ 
and $R_{max}$.The $t_{Bmax}$  estimates obtained by the two 
versions of the code almost coincide; the maximum 
in $B$  occurred about two days earlier than the maximum
in $V$, which is typical of SNe Ia. However, 
the values of $\Delta m_{15}$  found by the two versions of the 
SNOOPY code slightly differ, while the estimates of 
$E(B-V)_{host}$  and $\mu$  do not agree very well with those 
found by analyzing the $(B-V)$  color curves and from 
the redshift of the galaxy UGC 8255. Some differences
in the behavior of the light and color curves of 
SN 2009nr belonging to the SN 1991T subtype from 
most of the "normal" SNe Ia are probably responsible
for these discrepancies. Note that the absolute 
magnitude of SN 2009nr calculated with SNOOPY 
is $M_V = -19.6$,  which is rather close to our previous 
estimate, but SNOOPY adopts an excessively large 
extinction and, accordingly, a smaller distance. 

\medskip
{\bf DISCUSSION} 

\medskip
Our study of SN 2009nr reveals its great similarity 
to SN 1991T in both light-curve shape and luminosity.
Noticeable differences include the higher decline 
rate of SN 2009nr at the late stage, which manifests 
itself in both $U,V,R,I$  light curves and "quasibolometric"
light curves, and the differences in color 
curves. 

The luminosity of SN 2009nr is very high, but 
it falls nicely on the plot of absolute $V$  magnitude 
against decline rate presented in Hicken et al. (2009). 

SN 2009nr is located far from the center of the 
Scd galaxy UGC 8255. The projection of the distance
from the SN to the galaxy center onto the 
plane of the sky is 14.6 kpc and this is only the 
lower limit for the true distance. The galaxy radius
to the 25 $m_B/\square ''$  
isophote is $46''$, or 10.9 kpc 
(de Vaucouleurs et al. 1991); consequently, the relative
distance of the SN from the galaxy center is 
1.34. Wang et al. (2008) plotted the distribution 
of SNe with various decline rates $\Delta m_{15}$  in relative 
radial galactocentric distance. This plot has no objects
in a fairly wide neighborhood of the point occupied
by SN 2009nr. Thus, no SNe with such a 
high luminosity at such a large relative galactocentric
distance have been observed previously. This 
is of great importance for elucidating the nature of 
the 1991T-class SNe Ia. Our estimates of the $^{56}$Ni 
mass for SNe 1991T, 1998es, and 2009nr show that 
$M(^{56}{\rm Ni}) > 0.8 M_{\odot}$ 
(possibly, even more than 1 $M_{\odot}$) 
for all three objects. It is believed that during the 
explosion of white dwarfs with a mass equal to the 
Chandrasekhar limit, the amount of synthesized $^{56}$Ni 
lies within the range from 0.4 to 0.8 $M_{\odot}$ 
(see, e.g., 
Mazzali et al. 2001). It can be surmised that the 
outbursts of SNe similar to SN 2009nr are explained 
by the explosions of objects with a mass larger than 
the Chandrasekhar limit. As we see from the above 
arguments, SN 2009nr obviously belonged to the old 
halo population of the spiral galaxy UGC 8255; the 
same conclusion was reached by Khan et al. (2011). 
In this case, the explosion probably occurred after the 
merger of white dwarfs; this assumption can explain 
both the large mass and the long lifetime of the system 
before its outburst. 

\medskip
{\bf ACKNOWLEDGMENTS} 

\medskip
This work was supported by the Ministry of 
Science of the Russian Federation (State contract 
No. 02.740.11.0249), the "Dynasty" Foundation of 
noncommercial programs, and the Russian Foundation
for Basic Research (project No. 10-02-00249a). 

\medskip
{\bf REFERENCES} 

\medskip
\refw
W. D. Arnett, Astrophys. J. 253, 785 (1982).

\refw
P. Balanutsa and V. Lipunov, Central Bureau Electronic
 Telegrams No. 2111 (2010).

\refw 
M. S. Bessel, Publ. Astron. Soc. Pacif. 102, 1181 
(1990).

\refw 
C. R. Burns, M. Stritzinger, M. M. Phillips, et al., 
Astron. J. 141, 19 (2011).

\refw 
C. Chevalier and S. A. Ilovaisky, Astron. Astrophys. 
Suppl. Ser. 90, 225 (1991).

\refw 
T. S. Chonis and C.M.Gaskell, Astron.J. 135, 264 
(2008).

\refw 
A. Conley, M. Sullivan, E. Y. Hsiao, et al., Astrophys. 
J. 681, 482 (2008).

\refw  
A. Elmhamdi, D. Tsvetkov, I. J. Danziger, and A. Kordi,
Astrophys. J. 731, 129 (2011). 

\refw 
R. J. Foley and G. Esquerdo, Central Bureau Electronic
Telegrams No. 2112 (2010). 

\refw 
G. Garavini, G. Folatelli, A. Goobar, et al., Astron. 
J. 128, 387 (2004). 

\refw
B. T. Hayden, P. M. Garnavich, R. Kessler, et al., 
Astrophys. J. 712, 350 (2010). 

\refw
M. Hicken, P. Challis, S. Jha, et al., Astrophys. J. 700, 
331 (2009). 

\refw
S. Jha, R. P. Kirshner, P. Challis, et al., Astron. J. 131, 
527 (2006). 

\refw
R. Kessler, J.P.Bernstein, D.Cinabro, et al., Publ. 
Astron. Soc. Pacif. 121, 1028 (2009). 

\refw 
R. Khan, J. L. Prieto, G. Pojmanski, et al., Astrophys. 
J. 726, 106 (2011). 

\refw
A. Landolt, Astron. J. 97, 337 (1992). 

\refw
W. Li, S. B. Cenko, and A. V. Filippenko, Central 
Bureau Electronic Telegrams No. 2111 (2010). 

\refw
V. Lipunov, V. Kornilov, E. Gorbovskoy, et al., Adv. 
Astron. 2010, Article ID 349171 (2010). 

\refw
P. Lira, N. B. Suntzeff, M. M. Phillips, et al., Astron. 
J. 115, 234 (1998). 

\refw
P. A. Mazzali, K. Nomoto, E. Cappellaro, et al., Astrophys.
 J. 547, 988 (2001). 

\refw
M. M. Phillips, P. Lira, N. B. Suntzeff,et al., Astron. 
J. 118, 1766 (1999). 

\refw
J. L. Prieto, A.Rest, and N.B.Suntzeff,Astron. J. 
647, 501 (2006). 

\refw
A. Saha, F. Thim, G. A. Tammann, et al., Astrophys. 
J. Suppl. Ser. 165, 108 (2006). 

\refw
D. Schlegel, D. Finkbeiner, and M. Davis, Astrophys. 
J. 500, 525 (1998). 

\refw
V. Stanishev, A. Goobar, S. Benetti, et al., Astron. 
Astrophys. 469, 645 (2007). 

\refw
M. Stritzinger, P. A. Mazzali, J. Sollerman, and 
S. Benetti, Astron. Astrophys. 460, 793 (2006). 

\refw
D. Yu. Tsvetkov and N. N. Pavlyuk, Astron. Lett. 30, 
32 (2004). 

\refw
D. Y. Tsvetkov, V.P.Goranskij, and N. N. Pavlyuk, 
Perem. Zvezdy 28 (8) (2008). 

\refw
G. de Vaucouleurs, A. de Vaucouleurs, H. G. Corwin, 
et al., Third Reference Catalogue of Bright Galaxies
(Springer-Verlag, New York, 1991). 

\refw
B. Wang, X. Meng, X. Wang, and Z. Han, Chin. 
J. Astron. Astrophys. 8, 71 (2008). 

\end{document}